\documentclass[prd,twocolumn,groupaddress,amsmath,amssymb]{revtex4}
\usepackage[dvipdfmx]{graphicx}
\usepackage{dcolumn}
\usepackage{bm}
\usepackage[dvipdfmx]{color}

\usepackage{color}
\usepackage{braket}

\begin{document}
\title{Gluonic-excitation energies and Abelian dominance in SU(3) QCD}
\author{Hiroki~Ohata}
\affiliation{Yukawa Institute for Theoretical Physics, Kyoto University, Kyoto 606-8502, Japan}
\author{Hideo~Suganuma}
\affiliation{Department of Physics, Kyoto University, Kitashirakawaoiwake, Sakyo, Kyoto 606-8502, Japan}
\date{\today}
\begin{abstract}
We present the first study of the Abelian-projected gluonic-excitation energies for the static quark-antiquark (Q$\bar{\rm Q}$) system in SU(3) lattice QCD at the quenched level, using a $32^4$ lattice at $\beta = 6.0$. 
We investigate ground-state and three excited-state Q$\bar{\rm Q}$ potentials, 
using smeared link variables on the lattice. 
We find universal Abelian dominance for the quark confinement force of  
the excited-state Q$\bar{\rm Q}$ potentials as well as the ground-state potential.
Remarkably, in spite of the excitation phenomenon in QCD, 
we find Abelian dominance for the first gluonic-excitation energy of about 1~GeV 
at long distances in the maximally Abelian gauge.
On the other hand, no Abelian dominance is observed for 
higher gluonic-excitation energies even at long distances.
This suggests that there is some threshold between 1 and 2~GeV 
for the applicable excitation-energy region of Abelian dominance.
Also, we find that Abelian projection significantly reduces 
the short-distance $1/r$-like behavior in gluonic-excitation energies.
\end{abstract}

\maketitle

\section{Introduction}
Since quantum chromodynamics (QCD) has been established as the fundamental theory of the strong interaction, analytical derivation of quark confinement directly from QCD has been an open problem.
The difficulty originates from non-Abelian dynamics and nonperturbative features of QCD, which are largely different from QED.
%

In 1970's, Nambu, 't~Hooft, and Mandelstam proposed an interesting idea that quark confinement might be physically interpreted with the dual version of the superconductivity \cite{NtHM}.
In the dual-superconductor picture for the QCD vacuum, there takes place one-dimensional squeezing of the color-electric flux among (anti)quarks by the dual Meissner effect, as a result of condensation of color-magnetic monopoles.

As for the possible connection between QCD and the dual-superconductor theory, 't~Hooft proposed an interesting concept of Abelian gauge fixing, a partial gauge fixing which 
diagonalizes some gauge-dependent quantity \cite{tH81}.
In particular, in the maximally Abelian (MA) gauge 
\cite{KSW87,SY90,SNW94,AS99,BCGMP03,DIK04,SS14}, which is a special Abelian gauge, the off-diagonal gluon has a large effective mass of about 1~GeV \cite{AS99}, and Abelian dominance of quark confinement is observed in lattice QCD \cite{SY90,SNW94,DIK04,SS14}.
Then, infrared QCD in the MA gauge becomes an Abelian gauge theory including the color-magnetic monopoles, of which condensation leads to the dual superconductor \cite{DGL}.

For other nonperturbative QCD quantities such as spontaneous chiral-symmetry breaking, Abelian dominance is observed in lattice QCD \cite{MW95}. 
However, it is nontrivial whether Abelian dominance holds for excitation phenomena in QCD or not, because this Abelianization scheme is conjectured to be valid only for low energies and long distances. 
For instance, Abelian dominance is shown to be decreasing with larger momentum 
or smaller distance from the gluon propagator in the MA gauge in both SU(2) and SU(3) lattice QCD \cite{AS99,BCGMP03}, although the propagator itself is not physical observable.

Then, in this paper, we study Abelian dominance for excited-state inter-quark potentials and gluonic-excitation energies in the MA gauge in SU(3) color QCD at the quenched level.
Here, the excited-state potentials are important for the description of excitation phenomena of QCD \cite{JKM03,TS03,BP04}, and the gluonic-excitation energies are interesting physical observables appearing in hybrid hadrons \cite{MS15}.
They have been investigated in lattice QCD \cite{JKM03,TS03,BP04}, and 
the lattice results have been compared as stringy modes in the string picture of hadrons for the static quark-antiquark system. 
In fact, apart from the linear confinement part, 
the excited-state potential has $1/r$ part 
with a positive coefficient in long distances of $r \ge$ 2~fm, 
and this $1/r$ behavior can be a signal of the stringy mode, 
although the stringy behavior is significantly suppressed 
in shorter distances than 2~fm \cite{JKM03,BP04}.
The gluonic-excitation energies are defined by the differences between the ground-state and excited-state potentials, and the lowest gluonic-excitation energy takes a larger value than 
about 1~GeV 
both for static quark-antiquark (Q$\bar{\rm Q}$) and 3Q systems in lattice QCD \cite{JKM03,TS03}. This large gluonic-excitation energy explains success of the quark model \cite{TS03}.

The organization of this paper is as follows. 
In Sec.~II, we briefly review the Abelian projection in lattice QCD in the MA gauge. 
In Sec.~III, we present our calculation procedure for the ground- and excited-state potentials 
in the static Q$\bar{\rm Q}$ system.
In Sec.~IV, we show the lattice QCD result for the excited-state potentials 
and the gluonic-excitation energies. 
Section~V is devoted to summary and conclusion.

\section{Maximally Abelian gauge fixing and Abelian projection}
We perform SU(3) lattice QCD simulations at the quenched level with the standard plaquette action \cite{R12}.
In lattice QCD, the gauge variable is described as the link variable $U_{\mu}(s) \equiv e^{iagA_{\mu}(s)} \in {\rm SU(3)}$, with the gluon field $A_{\mu}(s) \in {\rm su(3)}$, QCD gauge coupling $g$ and the lattice spacing $a$.
In this paper, we use the lattice size of 
$L^3 \times L_t=32^4$ at $\beta \equiv 6/g^2 = 6.0$.

As for the lattice spacing $a$, we take $a \simeq$ 0.1022(5)~fm from Ref.~\cite{SS14}, 
where $a$ is determined to reproduce the string tension $\sigma \simeq$ 0.89~GeV/fm 
with large-number gauge configurations. 
Using the pseudo-heat-bath algorithm, we generate 150 gauge configurations which are taken every 500 sweeps after a thermalization of 5000 sweeps. 
For the estimate of the gluonic-excitation energies, it is found to be enough to use such number of configurations, although much higher statistics would be desired for more accurate analysis.

We perform MA gauge fixing by maximizing the norm
\begin{align}
  R_{\rm MA}[U_{\mu}(s)] \equiv& \sum_s \sum_{\mu=1}^4 {\rm tr} \left( U_{\mu}^\dag(s) \vec{H} U_{\mu}(s) \vec{H} \right) \notag \\
  =& \sum_{s} \sum_{\mu=1}^4 \left( 1 - \frac{1}{2}\sum_{i \neq j} \left| U_{ij} \right|^2 \right) \label{eq:Abelian}
\end{align}
under the SU(3) gauge transformations for each gauge configuration.
Here, $\vec{H} \equiv \left( T_3, T_8 \right)$ is the Cartan subalgebra of SU(3), i.e.,
$T_3 = \frac{1}{2} {\rm diag} \left( 1, -1, 0 \right)$ and $T_8 = \frac{1}{2\sqrt{3}} {\rm diag}\left( 1, 1, -2 \right)$.
When link variables are diagonal, one finds $R_{\rm MA} / (4 L^3L_t) = 1$.

We numerically perform MA gauge fixing using the over-relaxation method for rapid achievement in the maximization algorithm.
As for the stopping criterion, we stop the maximization algorithm when the deviation $\Delta R_{\rm MA} / (4 L^3L_t)$ becomes smaller than $10^{-5}$. 
The converged value $\langle R_{\rm MA} / (4 L^3L_t)\rangle$ is $0.7322(2)$ for 150 configurations, where the value in parentheses denotes the standard deviation. 
Judging from the small deviation, our procedure seems to escape bad local minima, and we expect that the Gribov ambiguity does not affect our results.

Finally, we extract the Abelian part of the link variable
\begin{equation}
  u_{\mu}(s) = \exp \left\{ i \theta_{\mu}^3(s) T_3 + i \theta_{\mu}^8(s) T_8 \right\} \in 
{\rm U}(1)_3 \times {\rm U}(1)_8
\end{equation}
from the link variable in the MA gauge, $U_{\mu}^{\rm MA}(s) \in {\rm SU(3)}$, by maximizing the norm 
\begin{equation}
R_{\rm Abel} \equiv {\rm Re \, tr} \left\{ U_{\mu}^{\rm MA}(s) u_{\mu}^\dag(s) \right\} \in \left[ -\frac{1}{2}, 1 \right],
\end{equation}
so that the distance between $u_{\mu}(s)$ and $U_{\mu}^{\rm MA}(s)$ becomes the smallest in the SU(3) manifold. 
We thus find ``microscopic Abelian dominance'', i.e., $\langle R_{\rm Abel} \rangle = 0.9027(1)$ as a local indicator \cite{IS99}. 
Of course, this does not necessarily mean ``macroscopic Abelian dominance'' 
in long distances, because, for instance, the quark confinement 
is judged by the exponential damping behavior in the Wilson loop \cite{R12}, 
and any component can be dominant if its damping is small. 
Actually, we will see a counter-example in the second gluonic-excitation energy.

The Abelian projection is defined 
by the replacement of the SU(3) link variable $U_\mu(s)$ 
by the Abelian part $u_\mu(s)$ for each gauge configuration, i.e., 
$O[U_\mu(s)] \rightarrow O[u_\mu(s)]$ for QCD operators.

\section{Lattice QCD calculation of excited-state inter-quark potentials}
In this section, we briefly mention the lattice QCD formalism 
to obtain the excited-state Q$\bar{\rm Q}$ potentials and our numerical procedure.

\subsection{Formalism}
We explain the variational and diagonalization method 
to calculate the ground- and excited-state potentials, 
originally reported in Ref.~\cite{L90}, in the same manner with Ref.~\cite{TS03}. 
We denote the QCD Hamiltonian $\hat{H}$ and the physical eigenstates $\ket{n} \, (n=0, 1, 2, \dots)$ for the static Q$\bar{\rm Q}$ system. 
As the eigenvalues $\hat{H}\ket{n} = V_n\ket{n}$, we define the $n$th excited-state potential $V_n$ with $V_0 \le V_1 \le V_2 \le \cdots$ in the static Q$\bar{\rm Q}$ system,
where the eigenvalues physically mean ground and excited potentials between the quark and the antiquark. 
For the simple notation, the ground state is often regarded as the ``$0$th excited state''. 

As sample states for the static Q$\bar{\rm Q}$ system, 
we prepare arbitrary given independent Q$\bar{\rm Q}$ states, 
$\ket{\Phi_k}\, (k = 0, 1, 2,\dots)$. 
In general, each state can be expressed with a linear combination of 
the eigenstates $\ket{n}$ as
\begin{equation}
\ket{\Phi_k} = c_0^{k}\ket{0} + c_1^{k}\ket{1} + c_2^{k}\ket{2} + \cdots .
\end{equation}

Let us consider time evolution of the sample states $\ket{\Phi_k}$ 
with the spatial locations of the quark and the antiquark being fixed. 
The Euclidean time evolution of the Q$\bar{\rm Q}$ state $\ket{\Phi_k(t)}$ 
is expressed with the operator $e^{-\hat{H}t}$, and 
we introduce the generalized Wilson loop 
\begin{align}
  W_T^{jk} \equiv& \braket{\Phi_j(T) | \Phi_k(0)} \\
=& \braket{\Phi_j | e^{-\hat{H}T} | \Phi_k} \notag \\
  =& \sum_{m=0}^{\infty} \sum_{n=0}^{\infty} \bar{c}_m^j c_n^k \braket{m |e^{-\hat{H}T} |n} \notag \\
  =& \sum_{n=0}^{\infty} \bar{c}_n^jc_n^k e^{-V_n T}.
\end{align}
Here, we define the matrix $C$ and the diagonal matrix $\Lambda_T$ by $C^{nk} = c_n^k,\, \Lambda_T^{mn} = e^{-V_nT}\delta^{mn}$, and rewrite the above relation as $W_T = C^\dag \Lambda_T C$. 
Note that $C$ is not a unitary matrix, and hence this relation does not mean the simple diagonalization by the unitary transformation.


In the lattice QCD simulations, 
we numerically calculate the generalized Wilson loop $W_T^{jk}$, 
and extract the potentials $V_0, V_1, V_2, \dots$ 
from $W_T$ and $W_{T+1}$ by using the formula
\begin{align}
  W_T^{-1}W_{T+1} =& \left( C^\dag \Lambda_T C \right)^{-1} C^\dag \Lambda_{T+1} C \notag \\
  =& C^{-1} {\rm diag} \left( e^{-V_0}, e^{-V_1}, \dots \right) C.
\end{align}
In fact, $e^{-V_0}, e^{-V_1}, e^{-V_2}, \dots$ can be obtained as the eigenvalues of the matrix $W_T^{-1}W_{T+1}$, i.e., $W_T^{-1}W_{T+1} \psi_n = e^{-V_n} \psi_n$, 
and they are also the solutions of the secular equation
\begin{equation}
  {\rm det} \left( W_T^{-1}W_{T+1} - t \right) = \prod_n 
\left( e^{-V_n} - t \right) = 0. \label{eq:vari}
\end{equation}

In the lattice QCD calculation, we use the above-mentioned method 
and extract low-lying excited-state potentials numerically 
for SU(3) QCD and Abelian-projected QCD, respectively.

\begin{figure*}[t]
\centering
\includegraphics[width=17.8cm,clip]{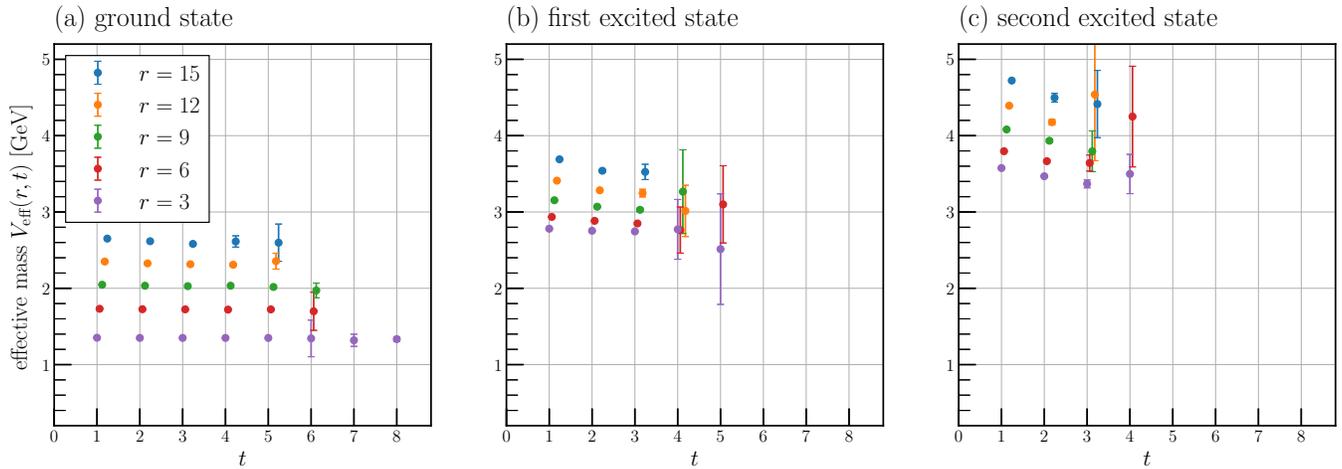}
\caption{
Effective mass plots $V_{\rm eff}(r,t)$ of the SU(3) potential $V(r)$ for 
(a) the ground state, (b) the first-excited state, and (c) the second-excited state 
in the static Q$\bar{\rm Q}$ system. 
Here, we display on-axis data of $r = 3, 6, 9, 12, 15$ in the lattice unit. Larger $r$ data are a bit shifted horizontally for visibility.
}
\label{Fig:effmass_MA}
\end{figure*}

\begin{figure*}[t]
\centering
\includegraphics[width=17.8cm,clip]{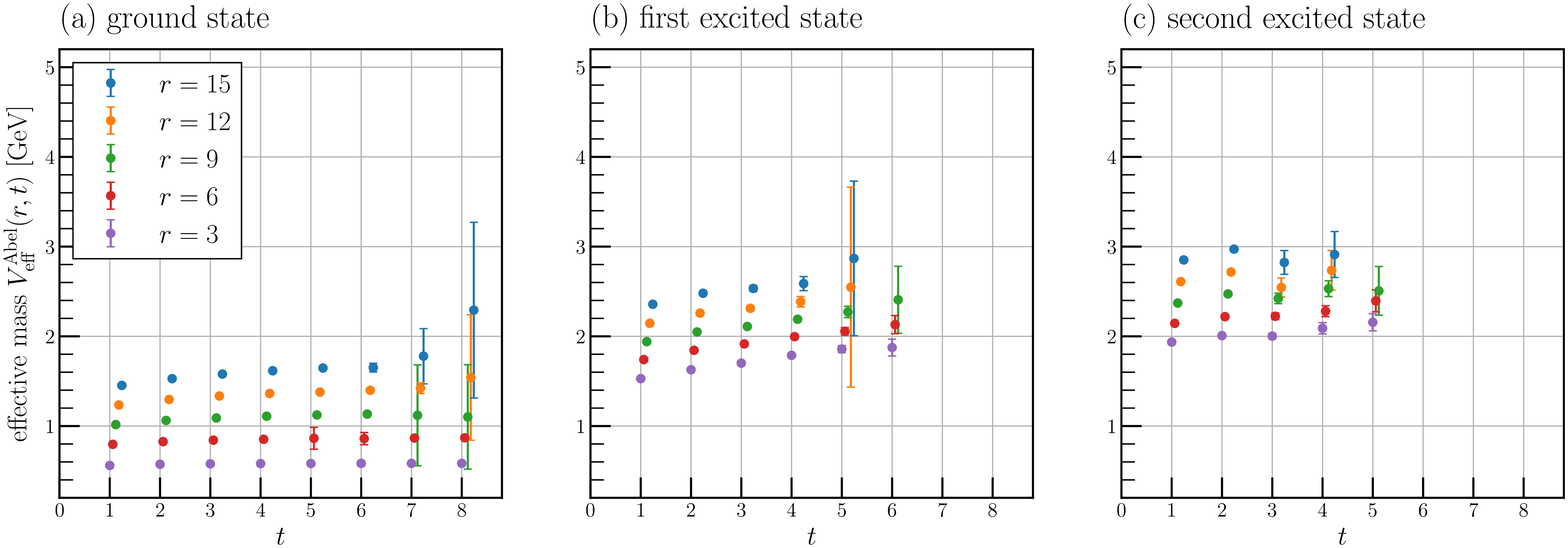}
\caption{
Effective mass plots $V_{\rm eff}^{\rm Abel}(r,t)$ 
of the Abelian potential $V^{\rm Abel}(r)$ for 
(a) the ground state, (b) the first-excited state, and (c) the second-excited state 
in the static Q$\bar{\rm Q}$ system. 
Here, we display on-axis data of $r = 3, 6, 9, 12, 15$ in the lattice unit. Larger $r$ data are a bit shifted horizontally for visibility.
}
\label{Fig:effmass_Abel}
\end{figure*}

\begin{figure*}[t]
\centering
\includegraphics[width=17.8cm,clip]{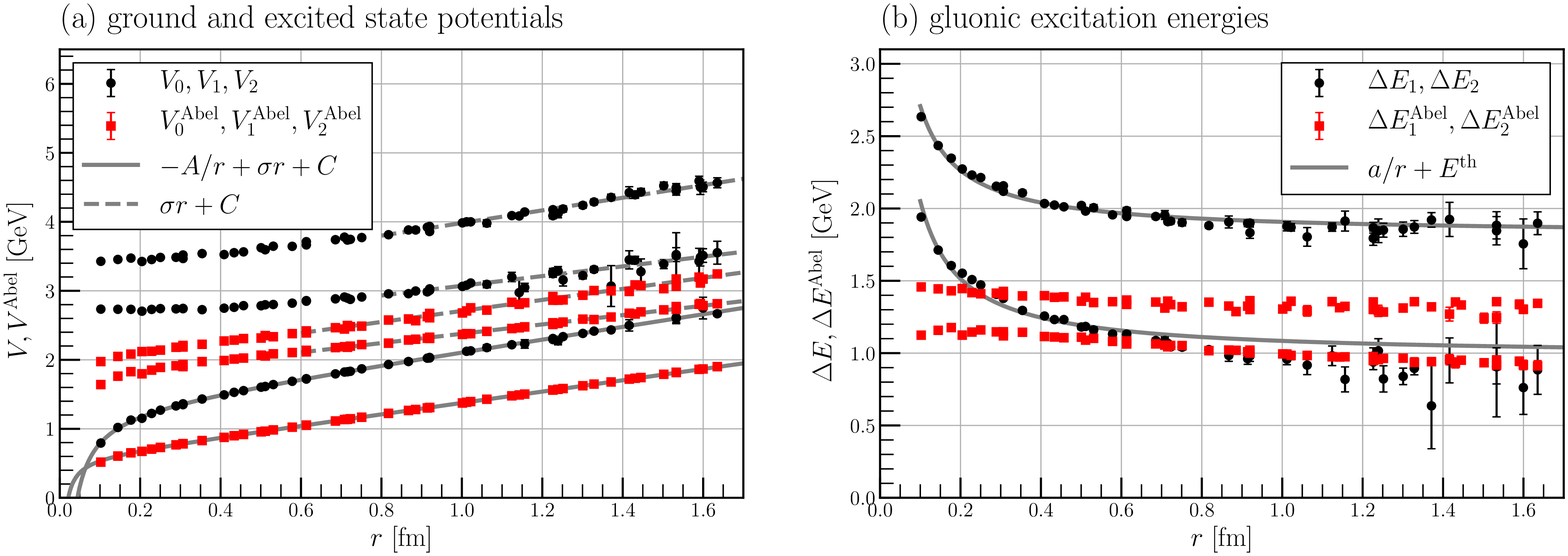}
\caption{
(a) The ground-state potential and two excited-state potentials 
in the static Q$\bar{\rm Q}$ system. The circles and the squares denote the SU(3) potentials $V_0, V_1, V_2$ and the Abelian potentials 
$V_0^{\rm Abel}, V_1^{\rm Abel}, V_2^{\rm Abel}$, respectively. 
The solid curves are the best fits with the Cornell potential $-A/r + \sigma r + C$ for $V_0$ and $V_0^{\rm Abel}$ for $r > 2a$.
The dashed lines are the best fits with $\sigma r+C$ for excited-state potentials 
$V_1$, $V_2$, $V_1^{\rm Abel}$, $V_2^{\rm Abel}$ at long distances. 
For the clear display, an irrelevant overall constant (+0.2~GeV) is added to $V_n^{\rm Abel}$.  
(b) Gluonic-excitation energies $\Delta E_n(r) \equiv V_n(r) - V_0(r)$. The circles and squares denote gluonic-excitation energies $\Delta E_1, \Delta E_2$ and the Abelian parts $\Delta E_1^{\rm Abel}, \Delta E_2^{\rm Abel}$, respectively. 
The curves are the best fits with the Ansatz $a/r + E^{\rm th}$ for SU(3) gluonic-excitations.
} 
\label{fig:pot}
\end{figure*}

\subsection{Numerical procedure}
In the practical calculation, we prepare four sample states $\ket{\Phi_k}$ for each gauge configuration by using the APE smearing method \cite{APE87}. 
Originally, the smearing method was developed as a useful technique to reduce the higher excitation components in a gauge-invariant manner. 
Here, the gauge-invariant Q$\bar{\rm Q}$ system consists of a quark, an antiquark, and gauge-covariant product of link variables connecting them, 
and, in the APE smearing method, gauge-covariant smeared link variables are used instead of original link variables \cite{APE87,TS02}.
We note that the smeared link variables depends on the iteration number 
$N_{\rm smr}$ of the smearing, and the $N_{\rm smr}$-times smeared states are generally 
linear independent each other when $N_{\rm smr}$ is different \cite{TS02}. 
Furthermore, the smeared states have only small components for highly excited states, 
and therefore they are appropriate as the sample states $\ket{\Phi_k}$ for 
the study of low-lying excitations \cite{TS03}.

Next, let us consider the quantum number of 
the Q$\bar{\rm Q}$ states obtained in this procedure with the APE smearing method, 
in terms of their parity (P), charge conjugation (C) and angular momentum ($\Lambda$), 
using the standard notation from the physics of diatomic molecules.
Here, the parity transformation means spatial inversion about the midpoint between 
the static quark and antiquark. 
The excitation modes obtained in this procedure 
are even ($g$) under charge-parity (CP) conjugation operation \cite{JKM03}, 
since the generalized Wilson loop $W_T$ with the reflection-symmetrically smeared states 
is invariant under CP transformation 
and all the CP-odd ($u$) components are cancelled in calculating the $W_T$.
%
%
For the CP-odd potentials, one needs to prepare CP-odd sample states, 
as is done in Ref.~\cite{JKM03}.
As for the total angular momentum ${\bf J}_g$ of gluons, 
the projection ${\bf J}_g \cdot \hat {\bf R}$ onto the molecular axis ${\bf R}$ 
gives a good quantum number and the magnitude of the eigenvalue of 
${\bf J}_g \cdot \hat {\bf R}$ is denoted by $\Lambda$ \cite{JKM03}.
In this procedure with the APE smearing method, only $\Sigma$ states with $\Lambda=0$ are expected to be obtained, since the smeared states are constructed 
in an axial-symmetric manner on the lattice. 
Also, there is a sign quantum number $\pm$ 
under a reflection in a plane containing the molecular axis \cite{JKM03}, 
and our axial-symmetric procedure makes only even ($+$) states.
In fact, in our procedure, we generate only CP-even ($g$), reflection-even ($+$) and $\Lambda=0$ states, which are denoted by the $\Sigma_g^+$ states \cite{JKM03} 
in the notation of the physics of diatomic molecules.

In the actual calculation, 
we prepare $8, 16, 24, 32$ times smeared states with the smearing parameter $\alpha = 2.3$, which is the standard value for the measurement of the SU(3) inter-quark potential \cite{TS02,BSS93}. 
Owing to the 8 times interval of the smearing, in most cases,  
the sample states $\ket{\Phi_k}$ seem to be practically linear independent 
with different patterns of the coefficients $c_n^k$, 
since the secular equation (\ref{eq:vari}) can be numerically solved. 
(If two of the sample states are not linearly independent, Eq.~(\ref{eq:vari}) cannot be solved.)
In this analysis, we make an assumption that higher excitation components $\ket{n}$ with $n \ge 4$ in the sample states $\ket{\Phi_k}$ are small enough and can be dropped off,
and solve the eigenvalue problem of the 4$\times$4 matrix $W_T^{-1}W_{T+1}$. 

As for the Abelian projection, 
we repeat just the same procedure by using Abelian link variables $u_\mu(s)$ instead of 
SU(3) link variables $U_\mu(s)$. 
Hereafter, we add the label ``Abel'' for the Abelian-projected physical quantities. 

In this way, we obtain the effective masses $V_{\rm eff}(r,t), V_{\rm eff}^{\rm Abel}(r,t)$ for the 0th, 1st, 2nd, 3rd excited state, respectively. 
The measurement is done for the on-axis and off-axis inter-quark directions as (1,0,0), (1,1,0), (2,1,0), (1,1,1), (2,1,1), and (2,2,1).
As the statistical error estimate, we adopt the jack-knife error estimate.


In calculating the potentials, higher excited states suffer larger systematic errors 
because the assumption of absence of higher modes becomes relatively more subtle.
Hence, we do not make quantitative analysis of the third-excited-state potentials, 
although preparing the four sample states definitely contributes to 
the significant error reduction for all the results.
To reduce the systematic errors further, we pick effective masses $V_{\rm eff}(r,t)$ at larger $t$ as long as the error is small. 

\section{Lattice QCD result}

In this section, we show the excited-state potentials and their Abelian projection 
in the static Q$\bar{\rm Q}$ system.

Figure~\ref{Fig:effmass_MA} shows the effective mass plots $V_{\rm eff}(r,t)$ 
for the SU(3) potentials $V_n(r)$ with $n=0,1,2$. 
Owing to the variational and diagonization method, for the low-lying states, 
$t$ dependence is small 
and an approximate plateau is observed even in small $t$ region, 
although higher excited state suffers larger statistical errors.
In this paper, we do not show the meaningless data 
with too large errors in figures. 
Here, we pick effective masses at $t=4, 3, 2$ as ground, first-excited, second-excited state SU(3) potentials, respectively.

Figure~\ref{Fig:effmass_Abel} shows the effective mass plots $V_{\rm eff}^{\rm Abel}(r,t)$ for the Abelian potentials $V_n^{\rm Abel}(r)$ with $n=0,1,2$.  
Compared with the SU(3) case, 
$V_{\rm eff}^{\rm Abel}(r,t)$ is slightly increasing as a function of $t$, 
and this might cause a systematic error of about 0.1~GeV on the choice of $t$. 
On the other hand, the statistical errors are smaller,  
because Abelian projection enhances the expectation value of the Wilson loop.
We pick effective masses at $t=4,3,2$ as ground, first-excited, second-excited state Abelian potentials, respectively.

Now, we show the ground, first-excited and second-excited state potentials 
$V_n \, (n=0,1,2)$
in the CP-even Q$\bar{\rm Q}$ system, and first and second gluonic-excitation energies $\Delta E_n \equiv V_n - V_0 \, (n=1,2)$ for both SU(3) and Abelian cases in Fig.~\ref{fig:pot}.

In the SU(3) case, the lattice results of $V_n \, (n=0,1,2)$ are consistent 
with those of three low-lying $\Sigma_g^+$ states in Refs.\cite{JKM03,BP04}, 
in terms of the overall behavior of $V_n(r)$, 
the infrared slope $\sigma_n \sim$1~GeV/fm of $V_n(r)$, 
and the interval $V_{n+1}-V_n \simeq$ 1~GeV at $r$=1~fm, 
except that the short-distance behavior of $V_2(r)$ 
is somehow different from Ref.\cite{BP04} because of the $0^{++}$ glueball mixture. 

As shown in Fig.~\ref{fig:pot} (a), all the SU(3) and Abelian potentials have 
approximately the same linear slope (the string tension) in long distances,  
which is also observed for the third-excited potential. 
This indicates universal Abelian dominance for the quark confinement force of 
the excited-state Q$\bar{\rm Q}$ potentials as well as the ground-state potential.

For more quantitative argument,  we evaluate the string tension $\sigma_n$ 
and the Abelian string tension $\sigma_n^{\rm Abel}$ 
from $V_n$ and $V_n^{\rm Abel}$, respectively, for $n=0,1,2$. 
For the ground-state potentials, $V_0$ and $V_0^{\rm Abel}$,  
we consider the best fits with the Cornell potential $-A/r + \sigma r + C$ 
(curves in Fig.~\ref{fig:pot}). From the fit for $ 2a < r \le 16a$, 
we evaluate the string tension $\sigma_0 \simeq$ 0.89(1)~GeV/fm and $A \simeq$ 0.291(9), 
and the Abelian-projected string tension $\sigma_0^{\rm Abel} \simeq$ 0.812(2)~GeV/fm 
and $A^{\rm Abel} \simeq$ 0.064(2).
(Reference~\cite{SS14} shows $\sigma^{\rm Abel} \simeq \sigma$ 
from the long-distance data with a large number of gauge configurations.)
For the excited-state potentials $V_n$, we evaluate 
the string tensions $\sigma_n$ from the fit with $\sigma r + C$ for large $r >$ 0.8~fm, 
and find $\sigma_1 \simeq$ 0.70(3)~GeV/fm and $\sigma_2 \simeq$ 0.92(3)~GeV/fm.
For the Abelian excited-state potentials $V_n^{\rm Abel}$, we evaluate 
the Abelian string tensions $\sigma_n^{\rm Abel}$ 
from the fit with $\sigma r + C$ for large $r >$ 0.6~fm, and find  
$\sigma_1^{\rm Abel} \simeq$ 0.676(7)~GeV/fm and 
$\sigma_2^{\rm Abel} \simeq$ 0.80(1)~GeV/fm.
Thus, we find $\sigma_n^{\rm Abel} \gtrsim 0.9 \sigma_n$ for $n=0,1,2$.

The gluonic-excitation energies are defined 
by the relative difference between the ground-state 
and excited-state potentials, $\Delta E_n(r) \equiv V_n(r) - V_0(r)$. 
Therefore their absolute values are physically meaningful,
while all potentials have ambiguity of an overall constant shift.
For the gluonic-excitation energies, we expect cancellation of systematic errors on $V_n$,  
especially for the Abelian part.

From Fig.~\ref{fig:pot}(b), 
the SU(3) gluonic-excitation energies $\Delta E_n(r)$ seem to be 
roughly approximated with the Ansatz $a_n/r + E_n^{\rm th}$ (the curves), and 
the best fit parameters are $a_1$= 0.54(2), $E_1^{\rm th}$ = 0.98(2)~GeV 
for $\Delta E_1(r)$, 
and $a_2$= 0.451(9), $E_2^{\rm th}$ = 1.818(7)~GeV for $\Delta E_2(r)$.
On the other hand, 
the Abelian-projected gluonic-excitation energies $\Delta E_n^{\rm Abel}(r)$ 
seem to be approximately constant: 
$\Delta E_1^{\rm Abel}(r) \simeq$1~GeV and $\Delta E_2^{\rm Abel}(r) \simeq$1.4~GeV. 
If $\Delta E_n^{\rm Abel}(r)$ is forced to be fit with the Ansatz 
$a_n^{\rm Abel}/r + E_n^{\rm th, Abel}$, the best fit parameters are 
$a_1^{\rm Abel}$= 0.14(2), $E_1^{\rm th, Abel}$ = 1.00(1)~GeV for $\Delta E_1^{\rm Abel}(r)$, 
and 
$a_2^{\rm Abel}$= 0.10(1), $E_2^{\rm th, Abel}$ = 1.323(6)~GeV for $\Delta E_2^{\rm Abel}(r)$.

Thus, we find three significant features 
for the gluonic-excitation energies $\Delta E_n(r)$ and $\Delta E_n^{\rm Abel}(r)$ 
as follows:
\begin{enumerate}
\item 
Abelian dominance is observed in the first gluonic-excitation energy 
in longer distances than about 0.7~fm: 
$\Delta E_1^{\rm Abel} \simeq \Delta E_1 \simeq$1~GeV.
In fact, a large gluonic-excitation energy of about 1~GeV is found 
even in Abelian-projected QCD.
\item 
No Abelian dominance is observed in the second gluonic-excitation energy, 
and the Abelian part is significantly smaller than the SU(3) result: 
$\Delta E_2^{\rm Abel} < \Delta E_2$. (This feature is also found for 
the third gluonic-excitation energy as $\Delta E_3^{\rm Abel} < \Delta E_3$.)
\item 
The short-distance $1/r$-like behavior is significantly reduced  
in the Abelian-projected gluonic-excitation energies $\Delta E_n^{\rm Abel}(r)$. 
\end{enumerate}

From the first two features, we conjecture that 
there is some threshold between 1 and 2~GeV
for the applicable excitation-energy region of Abelian dominance, 
and Abelian dominance holds below the threshold.
This seems to be qualitatively consistent with the behavior of 
the MA-gauge gluon propagator, which shows decreasing of Abelian dominance 
with larger momentum or smaller distance\cite{AS99,BCGMP03}. 

Here, Abelian dominance holds for nonperturbative properties such as confinement and spontaneous chiral-symmetry breaking, but does not hold for 
perturbative QCD.
Then, as an interesting conjecture, we expect that 
the first gluonic-excitation energy of about 1~GeV in long distances 
is nonperturbative, since it exhibits Abelian dominance.

On the other hand,
the higher gluonic-excitation energies, which do not show Abelian dominance, 
might have perturbative ingredients, which would obey 
$\Delta E^{\rm Abel}_{\rm pQCD} \simeq \frac{1}{4}\Delta E_{\rm pQCD}$,
according to the gluon-number reduction through Abelianization.

Finally, let us consider the significant reduction of 
the short-distance $1/r$-like behavior 
in the Abelian-projected gluonic-excitation energies.
If one considers the above-mentioned best fit with the Ansatz $a/r + E^{\rm th}$ for 
$\Delta E_n$ and $\Delta E_n^{\rm Abel}$ to be serious, 
one finds $a_n^{\rm Abel} \simeq \frac{1}{4} a_n$, 
which agrees with the gluon-number reduction through Abelianization, 
as is also seen in the perturbative one-gluon exchange.
Then, as an interesting possibility,  
the short-distance $1/r$-like behavior might originate from perturbative QCD, 
instead of nonperturbative QCD.
In any case, this finding would be a key to understanding  
the short-distance $1/r$ behavior 
in the excited SU(3) potentials for the static Q$\bar{\rm Q}$ system.

\section{Summary and Conclusion}
In this paper, we have presented the first study of 
the Abelian-projected gluonic-excitation energies in the static Q$\bar{\rm Q}$ system 
in SU(3) lattice QCD at the quenched level.
Using  smeared link variables on the lattice, 
we have examined four low-lying CP-even Q$\bar{\rm Q}$ potentials. 
We have found universal Abelian dominance for the quark confinement force 
also in the excited-state Q$\bar{\rm Q}$ potentials.

As a remarkable fact, 
we have found Abelian dominance in the first gluonic-excitation energy of about 1~GeV 
in long distances in the maximally Abelian gauge, although
it is an excitation phenomenon in QCD.
In contrast, no Abelian dominance has been observed 
in the second and higher gluonic-excitation energies.
From these two findings, we have conjectured that
there is some threshold for the applicable excitation-energy region 
of Abelian dominance between 1 and 2~GeV.

In addition, we have found that the short-distance $1/r$ behavior  
in gluonic-excitation energies is significantly reduced by the Abelian projection.
This finding would be a key to understand the short-distance $1/r$ behavior 
in the excited SU(3) potentials for the static Q$\bar{\rm Q}$ system.

As a future work, it is interesting to perform 
the similar study for the baryonic 3Q system. 
It is also meaningful to examine Abelian projection for 
CP-odd excited-state Q$\bar{\rm Q}$ potentials, 
using asymmetric sample states as in Ref.~\cite{JKM03}. 

It is also interesting to investigate the long-distance behavior 
of the gluonic-excitation energies in Abelian-projected QCD and 
to compare with the stringy mode of the Q$\bar{\rm Q}$ flux tube.
In SU(3) lattice QCD, the stringy modes grow up 
and appear in longer distances than 2~fm \cite{JKM03}.
Since Abelian-projected QCD also exhibits 
quark confinement and the flux-tube formation \cite{DIK04}, 
the stringy modes are expected also in the Abelian part in longer distances.

Also, it is meaningful to analyze our result in terms of low and high momentum gluon modes, 
since the gluon propagator shows that 
Abelian dominance decreases with larger momentum 
or smaller distance in the MA gauge in both SU(2) and SU(3) lattice QCD \cite{AS99,BCGMP03}.

\begin{acknowledgements}
H.S. is supported in part by the Grants-in-Aid for
Scientific Research [19K03869] from Japan Society for the Promotion of Science.
The lattice QCD calculations have been performed on NEC SX-ACE at Osaka University.
We have used LAPACK subroutines DGESVX and DGEEV to solve linear equations and eigenvalue problems, respectively. 
\end{acknowledgements}

\end{document}